\documentclass[11pt]{article}
\usepackage{amsmath,amssymb,epsfig,sint}

\newcommand{\R}{\rm I\kern-.2emR}
\newcommand{\C}{\rm \kern.25em\vrule height1.4ex
depth-.12ex width.06em\kern-.31em C}
\newcommand{\N}{{\rm I\kern-.16em N}}
\newcommand{\Z}{{\rm Z\kern-.35em Z}}

\newcommand{\rmd}{{\rm d}}
\newcommand{\rmO}{{\rm O}}

\newcommand{\be}{\begin{equation}}   
\newcommand{\ex}{\end{equation}}
\newcommand{\ba}{\begin{eqnarray}}
\newcommand{\ea}{\end{eqnarray}}

\newcounter{subequation}[equation]
\makeatletter

\expandafter\let\expandafter
\reset@font\csname reset@font\endcsname

\def\subeqnarray{\arraycolsep1pt
    \def\@eqnnum\stepcounter##1{\stepcounter{subequation}%
        {\reset@font\rm(\theequation\alph{subequation})}}
\jot5mm     \eqnarray}

\makeatother

\newcommand{\msbar}{{\rm \overline{MS\kern-0.14em}\kern0.14em}}

\renewcommand{\theequation}{\arabic{equation}}

\begin{document}


\begin{titlepage}

\begin{flushright}
   MPP-2009-10\\
   January 2009
\end{flushright}

\vskip 0.20 true cm

\begin{center}
{\Large\bf 
Logarithmic corrections to $\rmO(a^2)$ lattice artifacts} 
\end{center}
\vskip 1 true cm
\centerline{\large Janos Balog}
\vskip1ex
\centerline{Research Institute for Particle and Nuclear Physics}
\centerline{1525 Budapest 114, Pf. 49, Hungary}
\vskip 1 true cm
\centerline{\large Ferenc Niedermayer}
\vskip1ex
\centerline{Institute for Theoretical Physics, University of Bern}
\centerline{CH-3012 Bern, Switzerland}
\vskip 1 true cm
\centerline{\large Peter Weisz}
\vskip1ex
\centerline{Max-Planck-Institut f\"ur Physik}
\centerline{F\"ohringer Ring 6, D-80805 M\"unchen, Germany}
\vskip 1 true cm
\centerline{\bf Abstract}
\vskip 1.0ex
We compute logarithmic corrections to the $\rmO(a^2)$
lattice artifacts for a class of lattice actions for the non-linear
$\rmO(n)$ sigma-model in two dimensions. The generic leading artifacts
are of the form $a^2[\ln(a^2)]^{(n/(n-2))}$. We also compute
the next-to-leading corrections and show that for the case $n=3$
the resulting expressions describe well the lattice artifacts in the step 
scaling  function, which are in a large range of the cutoff apparently of 
the form $\rmO(a)$. An analogous computation should, if technically possible,
accompany any precision measurements in lattice QCD.

\vfill
\eject

\end{titlepage}



\noindent 1. Most of our knowledge concerning renormalization of quantum field theories
stems from perturbation theory. Although there are no rigorous proofs 
in general, many of the results are structural and 
hence considered to carry over to non-perturbative formulations. 
Indeed there is supporting evidence from various studies, e.g. of soluble 
models in 2 dimensions and of $1/n$ expansions of some theories.

The same situation holds concerning cutoff artifacts in lattice 
regularized theories. It is generally accepted that these artifacts
are summarized in Symanzik's effective action.
In this framework generic lattice artifacts are, in particular for
asymptotically free (or trivial) theories, expected to be 
integer powers in the lattice spacing $\rmO(a^p)\,,p=1,2,\dots$ up to 
possible multiplicative logarithmic corrections. This is an extremely 
important ansatz in the extrapolation of lattice data to the continuum 
limit, especially for present computations of lattice QCD
where the lattice spacings are typically around $0.1{\rm fm}$.

In this letter we will examine in more detail Symanzik's theory for the
2-dimensional non-linear $\rmO(n)$ $\sigma$-model. In such a simple
bosonic model lattice artifacts are expected to be of the form $\rmO(a^2)$. 
It was thus rather surprising that precision measurements \cite{HHNSW}, 
some years ago, of certain observables in the $\rmO(3)$ sigma model exhibited 
apparently linearly dependent $\rmO(a)$ artifacts
for a rather large range of computable lattice spacings.
The importance of finding the solution of this puzzle was emphasized by
Hasenfratz in his lattice plenary talk in 2001 \cite{PeterH}.

To define the measured quantity referred to above, one considers 
the model confined to a finite (1-dimensional) box of extension $L$ (with
periodic boundary conditions). The LWW coupling \cite{LWW} is defined as
\begin{equation}
u_0=L\,m(L)\,,
\end{equation}
where $m(L)$ is the mass gap of the theory in finite volume. 
Next one measures $u_1$, defined similarly with doubled box size.
In the continuum limit $u_1$ is a function of $u_0$, called the step
scaling function $u_1=\sigma(2,u_0)$. For the lattice regularized theory
there are lattice artifacts and
\begin{equation}
u_1=2L\,m(2L)=\Sigma(2,u_0,a/L)\,.
\end{equation}
The advantage of this measurement for the purpose of studying
lattice artifacts is that there is no need to 
know the box size $L$ or the mass gap $m(L)$ in physical units.

\begin{figure}
\begin{center}
\psfig{figure=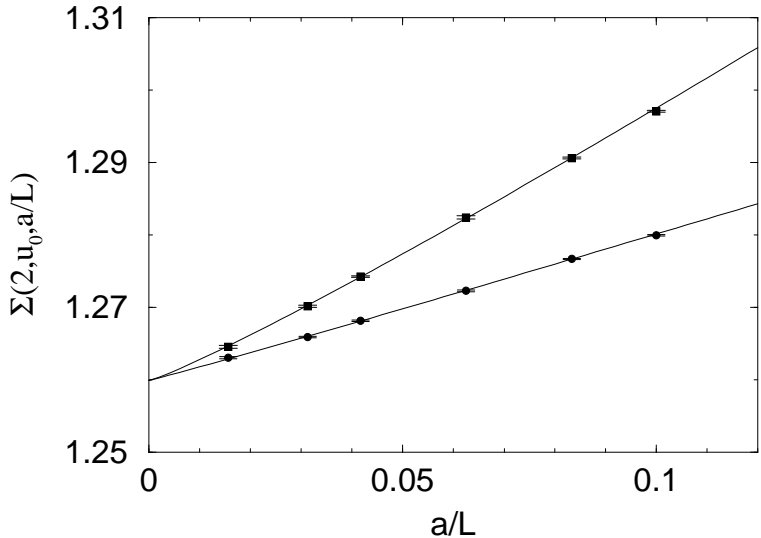,width=10cm}
\end{center}
\vspace{-0.5cm}
\caption{\footnotesize 
Monte Carlo measurements of the step scaling function at
$u_0=1.0595$. The data for the larger artifacts correspond to a modified
action. The fit contains $a$ and $a\ln a$ terms.
}
\label{lww}
\end{figure}

The results of the MC measurements are shown in Fig.~\ref{lww}. 
One can see that the lattice artifacts (cutoff effects) are very nearly 
linear as function of the lattice spacing $a$ both for the case of the 
standard lattice action (ST) and for a modified action (MOD). 
Although the effects are in this case relatively very small, 
they seem not of the theoretically expected form. 
Note however the encouraging feature that computations with different 
lattice actions are consistent with the same continuum limit, supporting 
the crucial concept of universality. 

\begin{figure}
\begin{center}
\psfig{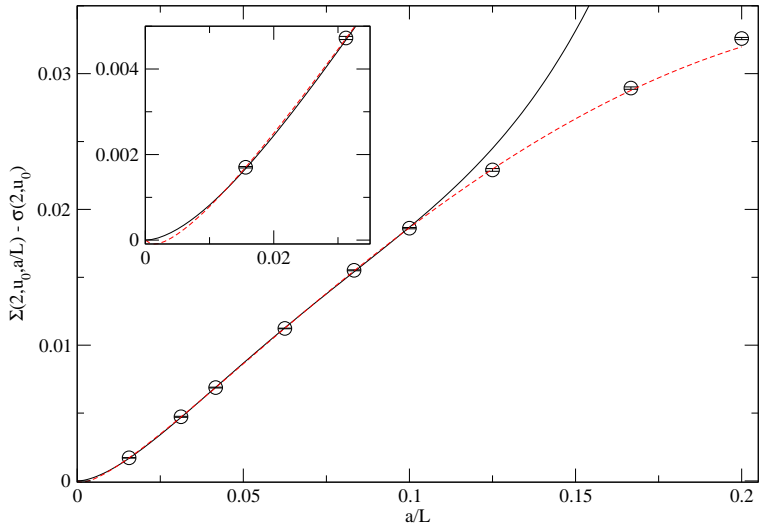}
\end{center}
\vspace{-0.5cm}
\caption{\footnotesize 
MC measurements of the step scaling function at $u_0=1.0595$. 
The curves are ``linear" (dashed line) and ``quadratic"
(solid line) fits described in the text.}
\label{lww2}
\end{figure}

The 2-dimensional O$(3)$ model is integrable and the finite volume mass gap
(and hence the step scaling function) is exactly calculable using 
thermodynamic Bethe Ansatz techniques~\cite{BH}. 
However, it turned out that the knowledge of the exact 
continuum limit did not help much in clarifying the problem of artifacts.
This is shown in Fig.~\ref{lww2}, where the exact continuum values are
already subtracted. The ``linear" fit assuming the functional form
$\delta(a)=c_1a+c_2a\ln a+c_3a^2$,
still gives a much better representation ($\chi^2/\mathrm{dof}=0.9$) 
than a ``quadratic" fit of the form 
$\delta(a)=c_1a^2+c_2a^2\ln a+c_3a^4$
which has an unacceptable $\chi^2/\mathrm{dof}=7.7$.
Note that although the $L/a\ge10$ data are used in the fits,
the ``linear" fit describes well also the three coarsest points.



In the early 80's Symanzik was working on the nature of lattice artifacts,
in particular with respect to his improvement program 
\cite{Sym,BMMS,SymCarg,SymBerl}.
In this letter the general theory is not discussed; we only consider 
Symanzik's theory applied to the 2-dimensional O$(n)$ $\sigma$-model. 
Nevertheless the spirit of the general theory
can already be understood by studying this example.
We shall see, based on Symanzik's theory,
why in the 2-dimensional $\sigma$-model quadratic artifacts are expected and 
how the particular logarithmic corrections predicted by 
this theory solves the puzzle of apparent linear artifacts. 
Similar conclusions have been reached previously in studies of the 
$\rmO(n)$ model in the first orders of the $1/n$ expansion 
\cite{KLW,BKLW}. We can here only give a brief 
description; full details will be presented in a separate paper \cite{BNW}.

\noindent 2. We write the lattice Lagrangian including the source terms symbolically as
\be
{\cal L}_{\rm latt}=\frac{1}{2\lambda_0^2}
\left(\partial_\mu S\cdot\partial_\mu S\right)^{\rm latt}-J\cdot S\,,
\qquad\qquad S^2=1\,,
\label{lattA2}
\end{equation}
with some lattice regularization of the kinetic term. This is used
in the generating functional for bare lattice
$S$-field correlation functions, which, after Fourier transformation,
become functions of the momenta, the bare lattice coupling $\lambda_0$
and the lattice spacing $a$. Performing, for fixed momenta and coupling,
a small $a$ expansion we can write them as
\be
{\cal G}^X_{\rm latt}(\lambda_0,a)={\cal G}^{X(0)}(\lambda_0,a)+
a^2\,{\cal G}^{X(1)}(\lambda_0,a)+\rmO\left(a^4\right)\,,
\label{cutoff}
\end{equation}
where the upper index $X$ symbolizes any $r$-point
correlation function (in $x$-space or in Fourier space), 
and both the scaling functions ${\cal G}^{X(0)}$ and the
leading cutoff corrections ${\cal G}^{X(1)}$ are still weakly
(logarithmically) depending on $a$.

The separation of the full lattice correlation function into a scaling
piece and cutoff corrections is unambiguous and straightforward in 
perturbation theory (PT).
In fact, in PT at $\ell$ loop order both terms are finite (order $\ell$) 
polynomials in $\ln a$. One of our main assumptions here is that the 
expansion (\ref{cutoff}) makes sense also beyond PT. 
Usual renormalization theory deals with the scaling part ${\cal G}^{X(0)}$.
Symanzik's important contribution was to show that 
the next term ${\cal G}^{X(1)}$ can be generated using an
effective Lagrangian.


\noindent 3. Symanzik's local effective Lagrangian ${\cal L}_{\rm eff}$ can be 
specified in the continuum in $D=2-\varepsilon$ dimensions in the 
framework of dimensional regularization:
\be
-{\cal L}_{\rm eff}=-{\cal L}+a^2\sum_{i=1}^7 Y_i(g,\varepsilon)U_i\,,
\label{LEL4}
\end{equation}
where ${\cal L}$ is the continuum Lagrangian with source terms
\be
{\cal L}=\frac{1}{2g_0^2}\left(\partial_\mu S\cdot
\partial_\mu S\right)-\frac{1}{g_0^2} I\cdot S\,,
\label{lcont}
\end{equation}
and the $U_i$ form a basis of local operators of dimension 4 invariant 
under the lattice symmetries, discussed below. 
There are no such operators of dimension 3 and hence no $\rmO(a)$ terms.

First we recall the continuum part (\ref{lcont}). Here $g_0$ is the bare 
coupling and 
the square of the bare O$(n)$ field $S^a(x)$ is normalized to unity
which is parameterized, as usual, by
$S^i=g_0\pi^i,\,\,i=1,\dots,n-1;\,\,\,
S^n=\sigma=\sqrt{1-g_0^2\pi^2}\,.$
The source dependent action ${\cal A}=\int\rmd^Dx{\cal L}(x)$ is used in 
the generating functional 
\be
{\cal Z}[I]=\int\left({\cal D}\pi\right)\,{\rm e}^{-{\cal A}},
\label{genfunc}
\end{equation}
which can be used to obtain bare correlation functions of the field $S^a(x)$:
\be
{\cal G}^{a_1\dots a_r}(x_1,\dots,x_r)=g_0^{2r}{\cal Z}^{-1}[I_0]
\frac{\delta}{\delta I^{a_1}(x_1)}\dots \frac{\delta}{\delta I^{a_r}(x_r)}
\Big\vert_{I_0}\,{\cal Z}[I]\,,
\label{corrfun}
\end{equation}
where the functional derivative is taken at
$I^a_0(x)=m_0^2\delta^{an}\,$;
i.e. a mass term (external magnetic field) is introduced to avoid infrared
singularities. As is well known, for O$(n)$ invariants the limit $m_0\to0$
can be taken at the end of the calculation.

In their seminal paper \cite{BZJG} Br\'ezin, Zinn-Justin and Le Guillou
prove the renormalizability of the O$(n)$ model using functional methods.
They showed that the generating functional ${\cal Z}^{-1}[I_0] 
{\cal Z}[I]$ is finite as a function of the renormalized  quantities
$j^a(x),g,\mu,m_R$ if we write
\be
\begin{split}
I^a(x)&=Z_1(g,\varepsilon)Z^{-1/2}(g,\varepsilon)g^2j^a(x)+I^a_0\,,\\
g_0^2&=\mu^\varepsilon Z_1(g,\varepsilon)g^2\,,\\
m_0^2&=Z_1(g,\varepsilon)Z^{-1/2}(g,\varepsilon)m_R^2\,,
\end{split}
\end{equation}
where in the minimal subtraction scheme the renormalization constants 
contain only pole terms in $\varepsilon$.

Functional derivation with respect to the source $j^a(x)$ gives renormalized
correlation functions, i.e. correlation functions of the renormalized fields
$S^a_R=Z^{-1/2}S^a$:
\be
\widehat{\cal G}^X_{(R)}(g,\mu,\varepsilon)=Z^{-r/2}(g,\varepsilon)
{\cal G}^X(g_0,\varepsilon)\,.
\end{equation}
We assume that
$X$ is O$(n)$ invariant and that the $m_R\to0$ limit has been 
taken. Finiteness means that the limit
\be
{\cal G}^X_{(R)}(g,\mu)=\lim_{\varepsilon\to0}
\widehat{\cal G}^X_{(R)}(g,\mu,\varepsilon)
\end{equation}
exists and defines the renormalized correlation function in two dimensions.

The renormalization group (RG) equations express the fact that the bare
correlation functions are independent of the renormalization scale $\mu$.
In terms of the renormalized correlation functions this is expressed as
\be
\left\{{\cal D}+\frac{r}{2}\gamma(g)\right\}{\cal G}^X_{(R)}(g,\mu)=0\,,
\label{RG1}
\end{equation}
where the RG differential operator is
\be
{\cal D}=\mu\frac{\partial}{\partial\mu}
+\beta(g)\frac{\partial}{\partial g}\,,
\end{equation}
and the RG beta and gamma functions are defined as
\be
\beta(g)=\frac{\varepsilon g}{2}-\frac{\varepsilon g}
{2+g\frac{\partial\ln Z_1(g,\varepsilon)}{\partial g}}=
-\beta_0g^3-\beta_1g^5-\beta_2g^7+\dots
\end{equation}
\be
\gamma(g)=\left\{\beta(g)-\frac{\varepsilon g}{2}\right\}
\frac{\partial\ln Z(g,\varepsilon)}{\partial g}=
\gamma_0g^2+\gamma_1g^4+\dots\,.
\end{equation}

Next, the dimension four operators $U_i$ in (\ref{LEL4}) are linear 
combinations of the 5 Lorentz scalar operators considered by
Br\'ezin et al \cite{BZJG}:
\begin{equation}
\begin{split}
{\cal O}_1&=\frac{1}{8}\left(\partial_\mu S\cdot\partial_\mu S\right)^2\,,
\quad\quad
{\cal O}_2=\frac{1}{8}\left(\partial_\mu S\cdot\partial_\nu S\right)
\left(\partial_\mu S\cdot\partial_\nu S\right)\,,\\
{\cal O}_3&=\frac{1}{2}\square S\cdot\square S\,,\quad\quad
{\cal O}_4=\frac{1}{2}\alpha
\partial_\mu S\cdot\partial_\mu S\,,\quad\quad
{\cal O}_5=\frac{1}{8}\,\alpha^2\,,
\end{split}
\label{obasis}
\end{equation}
where
\begin{equation}
\alpha=\frac{\square \sigma+I^n(x)}{\sigma}\,,
\end{equation}
and a further two operators which are only invariant under discrete 
lattice rotations:
\begin{equation}
A=\sum_{\mu=1}^D\widehat t_{\mu\mu\mu\mu}\,,\quad\quad
B=\sum_{\mu=1}^D\widehat k_{\mu\mu\mu\mu}\,,
\end{equation}
where $\widehat t$ and $\widehat k$ are the traceless parts
of the symmetric tensors $t,k$:
\be
\begin{split}
t_{\mu\nu\rho\sigma}&=
S\cdot\partial_\mu\partial_\nu\partial_\rho\partial_\sigma S\,,\\
k_{\mu\nu\rho\sigma}&=\frac{1}{3}\left\{
\left(\partial_\mu S\cdot\partial_\nu S\right)
\left(\partial_\rho S\cdot\partial_\sigma S\right)
+2\,\,{\rm perms}\right\}\,.
\end{split}
\end{equation}
Although the source dependent operators ${\cal O}_4$ and ${\cal O}_5$ 
look O$(n)$ non-invariant,
Br\'{e}zin et al show that they must be included in the operator 
renormalization scheme for consistency. 
The operators renormalize multiplicatively according to
\be
U_{i(R)}={\cal K}_{ij}(g,\varepsilon)\,U_j\,,
\end{equation}
where the matrix of renormalization constants is block diagonal, 
consisting of a $2\times2$ block for $i,j=6,7$ ($\propto A,B$), and
a $5\times5$ block for $i,j=1,2,3,4,5$.
The operator renormalization matrix is of the form
\be
{\cal K}_{ij}(g,\varepsilon)=\delta_{ij}-\frac{g^2}{\varepsilon}\,k_{ij}+
\frac{g^4}{2\varepsilon}\,\nu_{ij}^{(2)}+\frac{g^4}{2\varepsilon^2}
\left(k_{is}k_{sj}+2\beta_0k_{ij}\right)+\dots\,.
\end{equation}
Our one loop result $k_{ij}$ for the $5\times5$ 
sub-block agrees with that which is obtained from the computation of
Br\'{e}zin et al \cite{BZJG} in the basis (\ref{obasis}).

Renormalized correlation functions with one operator insertion are  
given by  
\be
{\cal G}^X_{i(R)}(g,\mu)=\lim_{\varepsilon\to0}
Z^{-r/2}(g,\varepsilon){\cal K}_{ij}(g,\varepsilon)
{\cal G}^X_j(g_0,\varepsilon)\,.
\end{equation}
They satisfy the RG equation
\be
\left\{{\cal D}+\frac{r}{2}\gamma(g)\right\}{\cal G}^X_{i(R)}(g,\mu)
+\nu_{ij}(g){\cal G}^X_{j(R)}(g,\mu)=0\,,
\end{equation}
where the anomalous dimension matrix is defined by
\be
\nu_{ij}(g)={\cal K}_{is}(g,\varepsilon)\left(\beta(g)-
\frac{\varepsilon g}{2}\right)\frac{\partial 
\left({\cal K}^{-1}\right)_{sj}(g,\varepsilon)}
{\partial g}=-k_{ij}g^2+\nu_{ij}^{(2)}g^4+\nu_{ij}^{(3)}g^6+\dots\,.
\end{equation}
We have chosen the basis $U_i$ such that the one-loop anomalous 
dimension matrix $k_{ij}$ is diagonal of the form
\be
k_{ij}=2\beta_0\Delta_i\delta_{ij}\,,
\end{equation}
with eigenvalues corresponding to
\be
\Delta_i=\left\{
\,\frac{n}{n-2}\,;\,-1\,;\,0\,;\,\frac{1-n}{n-2}\,;\,\frac{1}{n-2}\,;
\,0\,;\,-1\,\right\}\,.
\label{spec}
\end{equation}

Now defining
\be
\widehat{c}_j(g,\varepsilon)=\sum_{i=1}^7Y_i(g,\varepsilon)
{\cal K}^{-1}_{ij}(g,\varepsilon)
\end{equation}
we can rewrite the second term in (\ref{LEL4}) as
\be
\sum_{i=1}^7Y_i\,U_i=\sum_{i=1}^7\widehat{c}_i\,U_{i(R)}\,.
\end{equation}
The limit $\widehat c_i(g,0)$ must therefore exist:
\be
c_i(g)=\widehat c_i(g,0)=
\sum_{\ell=0}^\infty c_i^{(\ell)}g^{2\ell}\,.
\label{ci}
\end{equation}

The coefficients $c_i$ depend on the particular lattice action.
In our investigations we only considered actions quadratic in
the spins:
\be
{\cal A}=\frac{\beta}{2}\sum_{x,y}\sum_a S^a(x)K(x-y)S^a(y)\,,
\label{action}
\end{equation}
with $\beta=1/\lambda_0^2$. Here $K$ is short range, satisfying 
$\sum_x K(x)=0\,,$ and $K(z)=K(Rz)\,,$
where $R$ is a lattice rotation or reflection. 
For the standard action 
$K(z)=\sum_\mu \left[2\delta_{z,0}-\delta_{z,\hat{\mu}}
-\delta_{z,-\hat{\mu}}\right]\,.$

The tree level effective action for this class of lattice actions
involves only the operators $A$ and ${\cal O}_3$:
\be
-{\cal L}_{\rm eff}^{(0)}=-\frac{1}{2\lambda_0^2}\left(\partial_\mu S\cdot
\partial_\mu S\right)+\frac{a^2}{\lambda_0^2}\left\{
\frac{e_4}{24}A+
\frac{e_0}{16}{\cal O}_3\right\}+\rmO\left(a^4\right)\,,
\label{L0}
\end{equation}
where $e_0=e_4=1$ for the standard action, and off-shell improved actions
are such that $e_0=e_4=0$. 
From this we obtain directly the leading coefficients $c_i^{(0)}$ in (\ref{ci}). 

\noindent 4. With these preparations the precise relation between the lattice 
correlation functions and those obtained by using the effective action
can now be specified as
\be
\begin{split}
{\cal G}^{X(0)}(\lambda_0,a)&=
y^r(g){\cal G}^X_{(R)}\left(g,\frac{1}{a}\right)\,,\\
{\cal G}^{X(1)}(\lambda_0,a)&=y^r(g)\sum_{i=1}^7c_i(g)
{\cal G}^X_{i(R)}\left(g,\frac{1}{a}\right)\,,
\end{split}
\label{Sym01}
\end{equation}
where we have identified (for simplicity) the scale parameter $\mu$ of 
dimensional regularization with the inverse of the lattice spacing $a$. 
The finite wave function renormalization constant 
$y(g)$ comes from the relation $j^a(x)=y(g)\,J^a(x)$
between the lattice source $J$ and the (renormalized) 
dimensional regularization source $j$.

The result (\ref{Sym01}) can also be written as
\be
{\cal G}^X_{\rm latt}(\lambda_0,a)=
{\cal G}^{X(0)}(\lambda_0,a)\left\{
1+a^2\delta^X(\lambda_0,a)\right\}+\rmO\left(a^4\right),
\end{equation}
where
\be
\delta^X(\lambda_0,a)=\sum_{i=1}^7c_i(g)\,\delta^X_i(g,a)
\end{equation}
\be
\delta^X_i(g,a)=\frac{{\cal G}^X_{i(R)}\left(g,\frac{1}{a}\right)}
{{\cal G}^X_{(R)}\left(g,\frac{1}{a}\right)}\,.
\label{cuti}
\end{equation}
It is easy to see that the functions $\delta^X_i$ 
satisfy the RG equation
\be
\left\{-a\frac{\partial}{\partial a}+\beta(g)\frac{\partial}{\partial g}\right\}
\,\delta^X_i(g,a)=-\nu_{ij}(g)\,\delta^X_j(g,a)\,.
\label{rg}
\end{equation}
To derive an explicit expression for $\delta^X$ we must solve this 
partial differential equation. For this purpose we introduce the matrix 
$U_{ij}(g)$, which solves the ordinary differential equation
\be
U_{ij}^\prime(g)=-\rho_{is}(g)\,U_{sj}(g)\,,
\label{Uij}
\end{equation}
where
\be
\rho_{ij}(g):=\frac{\nu_{ij}(g)}{\beta(g)}=
\frac{2\Delta_i}{g}\,\delta_{ij}
+\sum_{\ell=2}^\infty\rho^{(\ell)}_{ij}\,g^{2\ell-3}.
\end{equation}
If we find the solution of (\ref{Uij}) we can write the general solution
of (\ref{rg}) as
\be
\delta^X_i(g,a)=U_{ij}(g)\,D^X_j(\Lambda)\,,
\end{equation}
and the lattice artifacts are of the form
\be
\delta^X(\lambda_0,a)=\sum_{i=1}^7\,\hat v_i(g)\,D^X_i(\Lambda)\,,
\quad\quad
\hat v_i(g)=\sum_{s=1}^7\,c_s(g)\,U_{si}(g)\,.
\label{arti1}
\end{equation}
The functions $D^X_j$ depend only on $\Lambda$, the RG invariant 
combination of $g$ and $a$. These functions are non-perturbative and 
depend on the quantity ($X$) we are considering. On the other hand, the
coefficients $\hat v_i(g)$ are perturbative and 
they remain the same for all physical quantities (but depend on the 
lattice action we started with).

We take the following ansatz: 
\be
U_{ij}(g)=\left\{\delta_{ij}
+\sum_{\ell=2}^\infty\,k^{(\ell)}_{ij}\,g^{2\ell-2}\right\}\,g^{-2\Delta_j}\,,
\end{equation}
Here the coefficients $k^{(\ell)}_{ij}$ may still weakly (logarithmically) 
depend on the coupling. This can arise if the difference between two
eigenvalues $\Delta_i-\Delta_j$ is a non-zero integer, which is
possible in our case, i.e. for $n=3$. We will however ignore this subtlety,
because we have verified that for the quantities we need here it plays no
role.

We can now write the lattice artifacts as
\be
\delta^X(\lambda_0,a)=\sum_{i=1}^7v_i(g)\,g^{-2\Delta_i}\,D^X_i(\Lambda)\,,
\label{arti2}
\end{equation}
where
\be
v_i(g)=c_i(g)+\sum_s c_s(g) 
\sum_{\ell=2}^\infty\,k^{(\ell)}_{si}\,g^{2\ell-2}
=\sum_{\ell=0}^\infty v_i^{(\ell)} \,g^{2\ell}\,.
\end{equation}
The spectrum of one-loop eigenvalues given by (\ref{spec}) plays a crucial
role in our considerations. The leading term corresponds to
\be
\Delta_1=\frac{n}{n-2}=n\chi=1+2\chi\,,
\end{equation}
(where $\chi:=1/(n-2)$) and the subleading one to
\be
\Delta_5=\frac{1}{n-2}=\chi\,.
\end{equation}
We thus have the leading expansion
\be
\begin{split}
\delta^X(\lambda_0,a)&=v_1\,D^X_1\,\left(g^{-2}\right)^{1+2\chi}+
v_5\,D^X_5\,\left(g^{-2}\right)^\chi+\dots,\\
&=D^X_1\,\left\{\left(g^{-2}\right)^{1+2\chi}v_1^{(0)}+
v_1^{(1)}\left(g^{-2}\right)^{2\chi}\right\}+
\rmO\left(\left(g^{-2}\right)^\chi\right)\,.
\end{split}
\label{arti3}
\end{equation}
It turns out that for the $n$=3 case $v^{(0)}_5=c^{(0)}_5=0$. 
This means that in this case 
the corrections start one power later and we have the leading expansion
\be
\delta^X(\lambda_0,a)=D^X_1\left\{
v^{(0)}_1\,g^{-6}+v^{(1)}_1\,g^{-4}+v_1^{(2)}\,g^{-2}\right\}
+\rmO(1)\,.
\end{equation}
The first expansion coefficients are
\be
\begin{split}
v^{(0)}_1&=c^{(0)}_1=\frac{e_0}{4(n-1)}\,,\\
v^{(1)}_1&=c^{(1)}_1+\sum_sc^{(0)}_sk^{(2)}_{s1}\,.
\end{split}
\end{equation}
Concretely we have
\be
k^{(2)}_{s1}=\frac{1}{2(\Delta_1-1-\Delta_s)}
\rho^{(2)}_{s1},
\end{equation}
which is different from zero for $s=1,2$ only.

This and the connection between the lattice coupling $\lambda_0$ and 
$g$ is all we need to write down the final result:
\be
\delta^X(\lambda_0,a)=\frac{e_0}{4(n-1)\left(2\pi\right)^{n\chi}}\,
D^X_1(\Lambda)\left\{
\left(\widetilde\beta\right)^{1+2\chi}+r^{(2)}
\left(\widetilde\beta\right)^{2\chi}\right\}+
\rmO\left(\widetilde\beta^\chi\right)
\label{veg}
\end{equation}
for $n\geq4$, and
\be
\delta^X(\lambda_0,a)=\frac{e_0}{(4\pi)^3}\,
D^X_1(\Lambda)\left\{
\widetilde\beta^3+r^{(2)}\widetilde\beta^2+r^{(3)}\widetilde\beta\right\}+
\rmO(1)\,,
\label{final3}
\end{equation}
for $n=3$, where we have introduced the inverse coupling 
$\widetilde\beta=2\pi/\lambda_0^2$. We expect this form of artifacts to
be generically present for all observables.
We note that our final result, eq. (\ref{veg}), 
is completely consistent with the large $n$ results of ref. \cite{KLW}.

We have computed the coefficient $r^{(2)}$ which is composed of three contributions:  
\be
r^{(2)}=r^{(2)}_{I}+r^{(2)}_{II}+r^{(2)}_{III}\,.
\label{r2pieces}
\end{equation}
For $n=3$ it would be nice to know the 3-loop coefficient 
$r^{(3)}$; its computation would however be a major undertaking
\footnote{It is built from (among other things) the two-loop 
coefficient $c^{(2)}_1$ 
appearing in the effective action and the three-loop anomalous dimension 
matrix elements $\nu^{(3)}_{ij}$.}.


For the computation of the first term in (\ref{r2pieces}) 
\be
r^{(2)}_{I}=\frac{8\pi(n-1)}{e_0}c^{(1)}_1\,,
\end{equation}
we need the 1-loop coefficients of the effective action.
We obtained these by
calculating the 2- and 4-point functions in both regularizations.
More precisely we need the scaling part and the O$(a^2)$
piece of the lattice correlation functions, and in the continuum we need 
the original correlation functions as well as the ones where those dimension
four operators that appear in the tree level effective action are inserted.
This is a long computation\footnote{In the case of operator insertions there
are many Feynman diagrams to be calculated. In the 4-point function case
for example there are 11 non-trivial diagrams, which cannot be reduced
to simpler ones like renormalization of tree diagrams or insertion of 
2-point function subgraphs with operator insertion. Each of these is
a complicated function of the four momenta and since they are topologically
distinct, it is difficult to automate the calculation.}, the details
of which will be published in ref.~\cite{BNW}. 

The second term in (\ref{r2pieces})
\begin{equation}
r^{(2)}_{II}=\frac{n}{n-2}(1-\psi)\,,
\end{equation}
involves the 1-loop relation between the 
lattice coupling $\lambda_0$ and the renormalized coupling 
of dimensional regularization, $g$:
\be
g^2=\lambda_0^2+\frac{\psi}{2\pi}\lambda_0^4+\dots
\end{equation}
This has been known for a wide class of actions for a long time.
e.g. for the case of ST at $n=3$ one gets $r^{(2)}_{II}=-3.98028$.

Finally the third term can be obtained from the $5\times5$ two-loop anomalous
dimension matrix of the dimensionally regularized scalar operators
\be
r^{(2)}_{III}=(2\pi)^2\left[
\frac{1}{n-2}\nu^{(2)}_{11}+\frac{1}{n}\nu^{(2)}_{21}\right]\,.
\end{equation}
This is again a lengthy computation which leads to the simple result
\begin{equation}
r^{(2)}_{III}=-2-\frac{9}{2(n-2)}\,.
\end{equation}



\noindent 5. For the standard action we finally obtain
\begin{equation}
r^{(2)}=-0.7625-\frac{5.6416}{n-2}-\frac{n}{4},
\end{equation}
giving a very large negative coefficient, $-7.1541$, for $n=3$.
Using the result (\ref{final3}), we can write in this case the
leading terms of the asymptotic series describing the lattice artifacts 
in terms of $a$ and the inverse coupling:
\begin{equation}
{\rm const.}\,a^2\,\left[\beta^3-1.1386\beta^2+\rmO(\beta)\right]\,.
\label{arti}
\end{equation}

\begin{figure}
\begin{center}
\psfig{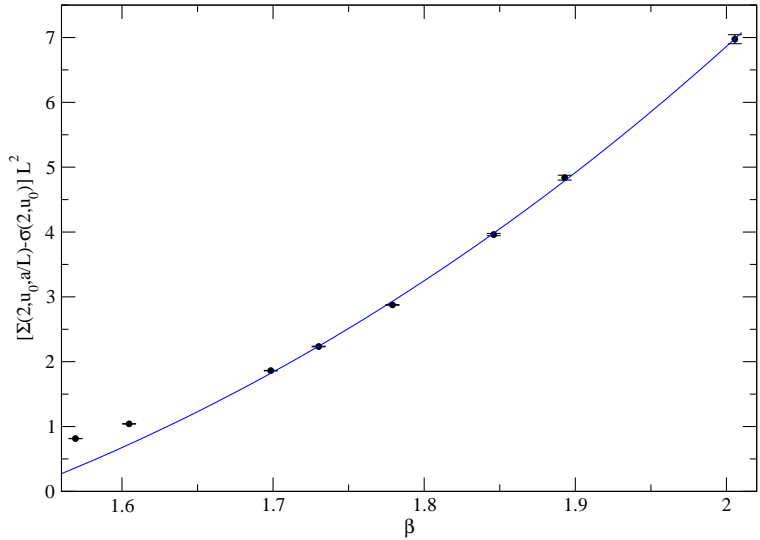}
\end{center}
\vspace{-0.5cm}
\caption{\footnotesize 
Modified 2-parameter fit: $3.06[\beta^3-1.14\beta^2-0.60\beta]$.
The points are from lattice sizes $L/a=5,6,10,12,16,24,32,64$;
the fit excludes the first two.
}
\label{fit2parfix}
\end{figure}

Due to the big negative value of the subleading term, 
the coupling dependent function in the square bracket
is a very rapidly growing function, which can compensate the decreasing
of one power of the lattice spacing, resulting in a behavior approximately
linear in $a$ in this limited range of coupling between $\beta\sim1.6$ and
$\beta\sim2.0$ as observed in \cite{HHNSW}. 
To demonstrate this, we multiplied the measured lattice
artifacts by $L^2$ to remove the $a^2$ factor and fitted a polynomial
cubic in $\beta$ to the resulting function. We first made a two-parameter
fit, with the result
$3.60\,\left[\beta^3-1.52\beta^2\right]\,.$
We also tried a 2-parameter fit, where the relative coefficient
of the subleading term was fixed at the value we obtained from
our analysis; we found
$3.06\left[\beta^3-1.14\beta^2-0.60\beta\right]\,.$
This fit is shown in Fig.~\ref{fit2parfix}. 
For both fits the two smallest lattices were omitted.
Both fits are good, indicating that the MC measurements
are consistent with the functional form given by our result (\ref{arti}).

For the modified action MOD, where in addition to the nearest neighbor
coupling there is also coupling between spins in the diagonal direction
(with equal strength), the results are similar. For this lattice
$e_0=5/3$ and
\begin{equation}
r^{(2)}=-3.0153-\frac{7.2625}{n-2}+\frac{3n}{20},
\end{equation}
giving $-9.8278$ for $n=3$. The artifacts are given by a formula
similar to (\ref{arti}), where the overall constant is bigger by
a factor $5/3$. For the relative (MOD/ST) artifacts
the overall non-perturbative
constant cancels and we find the following parameter-free 1-loop prediction
(for $n=3$):
\begin{equation}
{\rm MOD}/{\rm ST}=\frac{5}{3}\left[1+\frac{0.224}{\beta}
+\rmO(1/\beta^2)\right]\,,
\end{equation}
which varies between $1.85$ and $1.89$ in our limited range of coupling.
This is in very good agreement with the ratio of artifacts found in
\cite{HHNSW}, as can be seen in Fig.~\ref{lww}.

\noindent 6. We conclude that Symanzik's theory describes well lattice 
artifacts in the O(3) sigma-model. 
By looking into the details of this problem we had the opportunity to 
further gain confidence in Symanzik's theory of artifacts and improvement.
Similar computations should, in our opinion, accompany 
precision lattice studies of QCD in order to control and better
estimate systematic errors arising from lattice artifacts for 
extrapolations to the continuum limit.

{\it Acknowledgments}: J.~B. is grateful to the Max--Planck--Institut 
f\"ur Physik, where most
of this investigation was carried out, for its
hospitality. The authors would especially like to thank Peter Hasenfratz, 
who initiated this work and participated in the early stages of the 
project. This research was supported in part by the Hungarian
National Science Fund OTKA (under T049495).



\eject


\begin{thebibliography}{99}

\bibitem{HHNSW}
M.~Hasenbusch, P.~Hasenfratz, F.~Niedermayer, B.~Seefeld, 
U.~Wolff, \hfill\break
Nucl.\ Phys.\ Proc.\ Suppl.\  {\bf 106} (2002) 911.

\bibitem{PeterH}
P.~Hasenfratz,
Nucl. Phys. Proc. Suppl. {\bf 106} (2002) 159.

\bibitem{LWW}
M.~L\"{u}scher, P.~Weisz, U.~Wolff, 
Nucl. Phys. {\bf B359} (1991) 221.

\bibitem{BH}
J.~Balog, A.~Hegedus,
J.\ Phys.\ {\bf A37} (2004) 1881, 1903.

\bibitem{Sym}
K.~Symanzik,
Nucl. Phys. {\bf B226} (1983) 187; ibid 205.

\bibitem{BMMS}
B.~Berg, S.~Meyer, I.~Montvay, K.~Symanzik,
Phys.\ Lett.\ {\bf B126} (1983) 467.

\bibitem{SymCarg}
K.~Symanzik,
DESY79/76 (Carg\`ese lecture, 1979).

\bibitem{SymBerl}
K.~Symanzik,
{\it Some Topics In Quantum Field Theory,} C81-08-11-4.

\bibitem{KLW}
F.~Knechtli, B.~Leder, U.~Wolff, 
Nucl. Phys. {\bf B726} (2005) 421.

\bibitem{BKLW}
J.~Balog, F.~Knechtli, B.~Leder, U.~Wolff, 
PoS LAT205:253:2006.

\bibitem{BNW}
J.~Balog, F.~Niedermayer, P.~Weisz, in preparation

\bibitem{BZJG}
E.~Br\'{e}zin, J.~Zinn-Justin, J.~C.~Le Guillou, 
Phys. Rev. {\bf D14} (1976) 2615.




\end{thebibliography}
\end{document}